\documentclass[12pt]{article}
\usepackage{amssymb}
\input{psfig.tex}
\def\be{\nopagebreak[3]\begin{equation}}
\newcommand{\ee}{\end{equation}}
\def\ba{\begin{array}}
\def\ea{\end{array}}

\def\bea{\begin{eqnarray}}
\def\eea{\end{eqnarray}}
\begin{document}
\newcommand{\wb}{\bar}
\newcommand{\wh}{\widehat}
\newcommand{\wt}{\widetilde}
\newcommand{\MM}{{\cal M}}
\newcommand{\GG}{{\cal G}}
\newcommand{\AAA}{{\cal A}}
\newcommand{\II}{{\cal I}}
\newcommand{\DD}{{\cal D}}
\newcommand{\HH}{{\cal H}}
\newcommand{\DDA}{{\cal D}}
\newcommand{\BB}{{\cal B}}
\def\TH{{\cal T}}
\newcommand{\KK}{{\cal K}}
\newcommand{\p}{\partial}

\def\NCA{\em Nuovo Cimento}
\def\NIM{\em Nucl. Instrum. Methods}
\def\NIMA{{\em Nucl. Instrum. Methods} A}
\def\NPB{{\em Nucl. Phys.} B}
\def\PLB{{\em Phys. Lett.}  B}
\def\PRL{\em Phys. Rev. Lett.}
\def\PRD{{\em Phys. Rev.} D}
\def\ZPC{{\em Z. Phys.} C}

\def\st{\scriptstyle}
\def\sst{\scriptscriptstyle}
\def\mco{\multicolumn}
\def\epp{\epsilon^{\prime}}
\def\vep{\varepsilon}
\def\ra{\rightarrow}
\def\ppg{\pi^+\pi^-\gamma}
\def\vp{{\bf p}}
\def\ko{K^0}
\def\kb{\bar{K^0}}
\def\al{\alpha}
\def\ab{\bar{\alpha}}
\def\wt{\widetilde}
\newcommand{\NN}{{\cal N}}
\newcommand{\RR}{{\cal R}}
\newcommand{\refb}[1]{(\ref{#1})}
\def\be{\begin{equation}}
\def\ee{\end{equation}}
\def\bea{\begin{eqnarray}}
\def\eea{\end{eqnarray}}
\def\CPbar{\hbox{{\rm CP}\hskip-1.80em{/}}}

\renewcommand{\theequation}{\thesection.\arabic{equation}}

\title{ Superstring,Unifications and Dualities\footnote{IASSNS-I.H.P-Mars,99}\\
}
\author{
 Joseph Kouneiher,\\
 {\it U.M.R. 7596 (C.N.R.S,Universit{\'e} Diderot Paris7)} \\ 
{\it 2, place Jussieu 75005 Paris, France}\\
 kouneiher@paris7.jussieu.fr\\
 }

\date{}
\maketitle
\vskip.5cm

\abstract{
 I describe\footnote{This conference is the first part of two lectures treating the geometric principle lying behind superstring theory. It is an  introductory one.}our understanding of physics near the planck length, in particular the great progress of the last four years in string theory. Superstring theory, and a recent extension called $M$ theory, are leading candidates for a quantum theory that unifies gravity with the other forces. As such, they are certainly not ordinary quantum field theories. However, recent duality conjectures suggest that a more complete definition of these theories can be provided by the large $N$ limits of suitably chosen $U(N)$ gauge theories associated to the asymptotic boundary of spacetime.}

\section{Introduction}

Philosophically, one of the main achievement of quantum fields theory is to have constructed (in the transcendental sense) the third dynamical category of {\it{interaction}}\footnote { General relativity constructed the second dynamical category of causality i.e. the force. Indeed, in general relativity, the metric is no longer an a priori component but on the contrary a physical phenomenon which has to be determined. It is for this reason that metric can absorb the forces. So we have a conversion of the kinematical moment concerning metric into the dynamical moment concerning forces, a shifts from the metrical and global level to the local and differentiable one.}, Weyl's gauge principle converting gauge invariances into dynamical principles. According to quantum field theories the constitutive principles (relativity, symmetry) provides lagrangians, which in turn provide Feynman's integrals, wich provide themselves the models.\\
Gauges theories have shown that if we localisze the internal symmetries and if we impose the invariance of the Lagrangians for these supplementary symmetries, we can reconstruct in a purely mathematical manner the interacions Lagrangians. The mathematical constraints are so strong (renormalizability, elimination of anomalies, Higgs mechanism of symmetry breaking for confering a mass to the gauge bosons, etc.) that it is possible to infer from very few empirical data to the choice of a symmetry group. In superstring theory this fact is even more evident.\\\\
In addition, as we all know, quantum field theory has been extremely
successful in providing a description of elementary particles and
their interactions. However, it does not work so well for
gravity. If we naively try to quantize general relativity
(which is a classical field theory) using the methods of quantum
field theory, we run into divergences which cannot be removed by
using the conventional renormalization techniques of quantum
field theory. String theory is an attempt to solve this problem\cite{GSW}.

\section{A Review of Perturbative String Theory}
\label{spert}
\begin{figure}[!ht]
\begin{center}
\leavevmode
\psfig{file=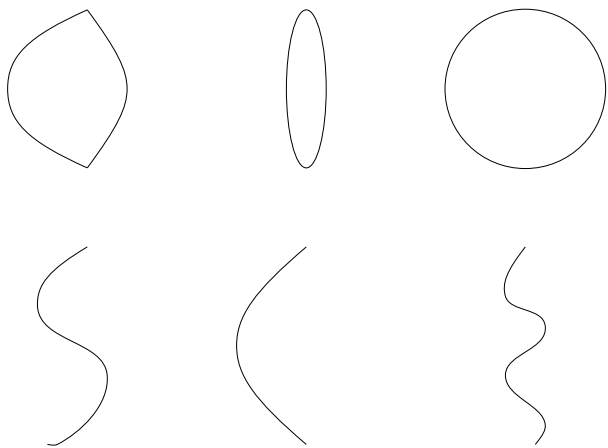,width=8cm}
\end{center}
\caption{Some vibrational modes of closed and open strings}
\label{f2}
\end{figure}

The basic idea in string theory is quite simple. According to
string theory, different elementary particles, instead of being
point like objects, are different vibrational modes of a string.
Fig.\ref{f2} shows some of the oscillation modes of closed strings
and open strings. 

The energy per unit length of the string, known as string
tension, is parametrized as $(2\pi\alpha')^{-1}$, where $\alpha'$
has the dimension of (length)$^2$. 
This theory automatically contains gravitational interaction
between elmentary particles, but in order to correctly reproduce
the strength of this interaction, we need to choose
$\sqrt{\alpha'}$ to be of the order of $10^{-33}cm$. Since
$\sqrt{\alpha'}$ is the only length parameter in the theory, the
typical size of a string is of the order of $\sqrt{\alpha'}\sim
10^{-33}cm$
$-$ a distance that cannot be resolved by present
day experiments. Thus there is no direct way of testing string
theory, and its appeal lies in its theoretical consistency.

\begin{figure}[!ht] 
\begin{center}
\leavevmode
\psfig{file=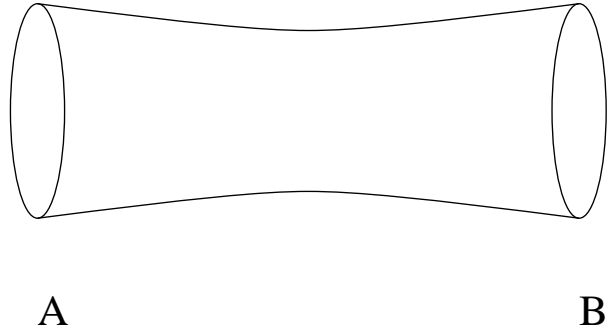,width=8cm}
\end{center}
\caption[]{\small Propagation of a closed string.} 
\label{f3}
\end{figure}

The basic principle behind constructing a quantum theory of
relativistic string is quite simple. Consider propagation of a
string from a space-time configuration A to a space-time
configuration B. During this motion the string sweeps out a two
dimensional surface in space-time, known as the string
world-sheet (see Fig.\ref{f3}). 
The amplitude for the
propagation of the string from the space-time position A to space-time
position B is given by the weighted sum over all world-sheet
bounded by the initial and the final locations of the string. The
weight factor is given by $e^{-S}$ where $S$ is the product of
the string tension and the area of the world-sheet. Let $\sigma$ be a parametrization of the string. If $\tau$ is its proper time, the parametrization of its world leaf is $X_\mu (\sigma , \tau)$ endowed with the metric $g_{ab}= g_{\mu \nu} \partial_{a}X^\mu \partial_{b}X^{\nu} (a,b= \sigma or \tau )$. This leads to the introduction of new Lagrangians, for instance the Polyakov Lagrangian:
\begin{equation}
L= \sqrt {g} g^{ab} \partial_{a}X^\mu \partial_{b}X^{\nu}
\end{equation}

with $g= \mid det (g_{ab})\mid$. One has to compute functionnal integrals of the following type:
\begin{equation}
Z= \sum_{topologies} \int_{metrics} Dg_{ab} \int_{leaves} DX^\mu  esp(i\frac{S}{\hbar})
\end{equation}

It turns out
that this procedure by itself does not give rise to a fully
consistent string theory. In order to get a fully consistent
string theory we need to add some internal fermionic degrees of
freedom to the string and generalize the notion of area by adding
new terms involving these fermionic degrees of freedom. This leads
to five (apparently) different consistent string theories in
(9+1) dimensional space-time.

\begin{figure}[!ht] 
\begin{center}
\leavevmode
\psfig{file=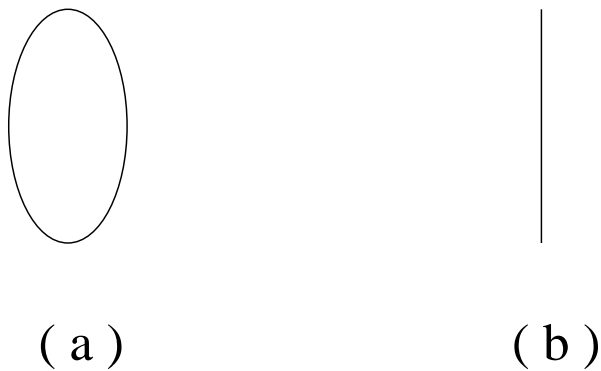,width=8cm}
\end{center}
\caption[]{\small (a) A closed string, and (b) an open string.} 
\label{f4}
\end{figure}

In the first quantized formalism, the dynamics of a point
particle
is described by quantum mechanics. Generalizing this we see that
the first quantized description of a string will involve a (1+1)
dimensional quantum field theory. However unlike a conventional
quantum field theory where the spatial directions have infinite
extent, here the spatial direction, which labels the coordinate
on the string, has finite extent. It represents a compact circle if
the string is closed (Fig.\ref{f4}(a)) and a finite line interval
if the string is open (Fig.\ref{f4}(b)). 

This (1+1) dimensional
field theory is known as the world-sheet theory.
The fields in this (1+1) dimensional
quantum field theory and the boundary conditions on
these fields vary in different string theories. 
Since the spatial direction of the
world-sheet theory has finite extent, each world-sheet field can
be regarded as a collection of infinite number of harmonic
oscillators labelled by the quantized momentum along this spatial
direction. Different states of the string are obtained by acting
on the Fock vacuum by these oscillators. This gives an infinite
tower of states.  Typically each string theory
contains a set of massless states and an infinite tower of
massive states. 
The massive string states typically have mass of the
order of $(10^{-33}cm)^{-1}\sim 10^{19}GeV$ and are far beyond
the reach of the present day accelerators. Thus the interesting
part of the theory is the one involving the massless states. We
shall now briefly describe the  interaction in
various string theories and their compactifications.

\subsection{Interactions}
\begin{figure}[!ht] 
\begin{center}
\leavevmode
\psfig{file=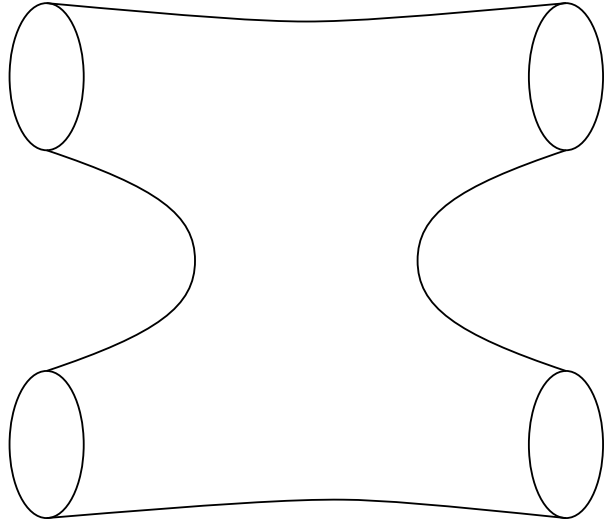,width=8cm}
\end{center}
\caption[]{\small A string world-sheet bounded by four external
strings.}
\label{f5}
\end{figure}

To  describe the theory we must also describe the
interaction between various particles in the spectrum (in string theories). In
particular, we would like to know how to compute a scattering
amplitude involving various string states. It turns out that
there is a unique way of introducing interaction in string
theory. Consider for example a scattering involving four external
strings, situated along some specific curves in space-time. 
The prescription for computing the scattering amplitude is
to compute the weighted sum over all possible string world-sheet
bounded by the four strings with weight factor $e^{-S}$, $S$
being the
string tension multiplied by the
generalized area of this surface (taking into account the fermionic 
degrees of freedom of the world-sheet). One such surface is shown
in Fig.\ref{f5}.

\begin{figure}[!ht] 
\begin{center}
\leavevmode
\psfig{file=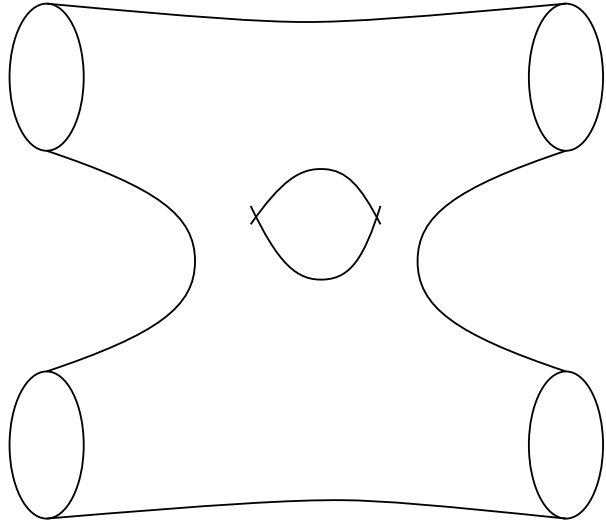,width=8cm}
\end{center}
\caption[]{\small A more complicated string world-sheet.} 
\label{f6}
\end{figure}

 If we imagine the time axis running from left
to right, then this diagram represents two strings joining into
one string and then splitting into two strings, $-$ the analog of
a tree diagram in field theory. A more complicated surface is
shown in Fig.\ref{f6}. 
This represents two strings joining into
one string, which then splits into two and joins again, and
finally splits into two strings. This is the analog of a one loop
diagram in field theory. The relative normalization between the
contributions from these two diagrams is not determined by any
consistency requirement. This introduces an arbitrary parameter
in string theory, known as the string coupling constant. However,
once the relative normalization between these two diagrams is
fixed, the relative normalization between all other diagrams is
fixed due to various consistency requirement. Thus besides the
dimensionful parameter $\alpha'$,  string theory
has a single dimensionless coupling constant.

So Feynman's interaction graphs are substituted by Riemann surfaces (which are topological configurations of interactions). For doing that, we need Riemann surfaces theory. For exemple we need Teichm\"{u}ler theory of moduli spaces for knowning exactly what are the automorphismes of a Riemann surface (what are its diffeomorphismes which are not isotopic to the identity, what are the complex structures compatible with a given differentiable structure, etc.). We need also the solution of the schottky problem. Let $S$ be a Riemann surface of genus $g$. It is well known that it is possible to find a basis $(a_{i}, b_{i}), i=1,...,g$ of the homology of $S$ and a basis $(\omega_{i}), j=1,..., g$ of the space of differentiable $1-$formes which are the simplest possible, that is to say which satisfy: 
\begin{equation}
\int_{a_{i}}\omega_{i} = \delta_{ij}
\end{equation}
and
\begin{equation}
\int_{b_{i}}\omega_{j} = \Omega_{ij}
\end{equation}
the matrix $\Omega = (\Omega_{ij})$ of periods being symmetric and of imaginary part positive definite: $Im\Omega > 0$. But if $g> 3$, the space of the matrices $\Omega$ which are symmetric and of imaginary part $> 0$ has a dimension $\frac{1}{2} g(g+1)$ which is greater than the dimension $3g-3$ of the moduli space of $S$. Therefore we must characterize the $\Omega$ which can be the period matrices of Riemann surfaces. This is the Schottky problem. It has been solved only in $1984$.

\subsection{Compactification}

The five different string theories mentioned above,
all live in ten space-time dimensions. Since our world is (3+1)
dimensional, these are not realistic string theories. However one
can construct string theories in lower dimensions using the idea
of compactification. The idea is to take the (9+1) dimensional
space-time
as the product of a $(9-d)$ dimensional compact manifold $\MM$
with euclidean signature
and a $(d+1)$ dimensional Minkowski space
$R^{d,1}$. Then, in the limit
when the size of the compact manifold is sufficiently small so
that the present day experiments cannot resolve this distance,
the world will effectively appear to be $(d+1)$ dimensional.
Choosing $d=3$ will give us a (3+1) dimensional theory.
Of course we cannot choose any arbitrary manifold $\MM$ for
this purpose; it must
satisfy the equations of motion of the effective field
theory that comes out of string theory.
One also normally considers only those manifolds which preserve
part of the space-time supersymmetry of the original ten
dimensional theory, since this guarantees vanishing of the
cosmological constant, and hence consistency of the corresponding
string theory order by order in perturbation theory.
There are many known examples of manifolds satisfying these
restrictions {\it e.g.} tori of different 
dimensions, K3, Calabi-Yau manifolds etc. For instance -using a Kaluza-Klein device - we can compactify $16$ dimensions (starting from $D= 26$) using the lattice of the roots of the Lie gauge group $E_{8}\otimes E_{8}$ and then compactify again $6$ dimensions :$M^{10} \rightarrow M^{4}\times K^{6}$. Physical constraints of preservation of supersymmetry impose for exemple that there exists on $K$ a spinor field $\xi$ which is constant for the covariant derivation (i.e. $D_{i}\xi = 0$).This fact imposes drastic constraints upon the geometry of $K^{6}$: the Ricci cruvature must be = 0, the holonomy group must be = $SU(3)$, the first Chern class $c_{1}(K)$ of $K$ must be = 0, there must be exist a k\"{a}hler metric on $K$, etc. ( In fact, according to a celebrated theorem of Calabi and Yau, a K\"{a}hler manifold $K^{2n}$ with $c_{1}(K)=0$ admits necessarily a k\"{a}hler metric with holonomy group $SU(n)$ (and not $O(2n)$)).\\
Instead of going via the effective action, one can also directly
describe these compactified theories as string theories. For this
one needs to
modify the string world-sheet action in such a way that it
describes string propagation in the new manifold $\MM\times R^{d,1}$, instead of in
flat ten dimensional space-time. This modifies the world-sheet
theory to an interacting non-linear $\sigma$-model instead of a
free field theory. Consistency of string theory puts
restriction on the kind of manifold on which the string can propagate.
At the end both approaches yield identical results.

 The effect of this compactification
is to periodically identify some
of the bosonic fields in the string world-sheet field
theory $-$ the fields which represent
coordinates tangential to the
compact circles. One effect of this is that the momentum carried
by any string state along any of these circles is quantized in
units of $1/R$ where $R$ is the radius of the circle. But that is
another novel effect: we now have new states that correspond to
strings wrapped around a compact circle. For such a states, as we
go once around the string, we also go once around the compact
circle. These states are known as winding states and play a
crucial role in the analysis of duality symmetries.
\begin{figure}[!ht]
\begin{center}
\leavevmode
\psfig{file=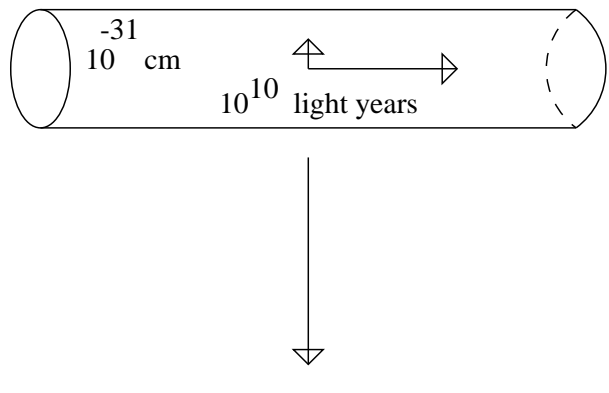,width=8cm}
\end{center}
\caption{Example of Compactification. The two dimensional
cylinder appears to be one dimensional if the radius is very
small.}
\label{f7}
\end{figure}

It turns out that there are many different choices for this six
dimensional compact manifold (case d = 3). Thus each of the five string
theories in (9+1) dimensions gives rise to many possible string
theories in (3+1) dimensions after compactification. Some of
these theories come tantalizingly close to the observed universe.
In particular one can construct models with:
\begin{itemize}
\item[i)] Gauge group containing the standard model gauge group
$SU(3)\times SU(2)\times U(1)$,
\item[ii)] Chiral fermions representing
three generations of quarks and leptons,
\item[iii)] N=1 supersymmetry,
\item[iv)] Gravity.
\end{itemize}
Furthermore unlike conventional quantum field theories which are
ultraviolet divergent but renormalizable, and quantum general
relativity which is ultraviolet divergent and not renormalizable,
string theories have no ultraviolet divergence at all!
\section{Unification}
The motivation for supersymmetry comes from  the idea of gauge
 unification. Recent experiments have yielded precise determinations
 of the strengths of the $SU(3)\times SU(2)\times U(1)$ gauges interactions - the
 analogs of the structure constant for these interactions. In quantum
 field theory theses values depend on the energy at which they are
 measured in  a way that depends on the particle content of the
 theory. Using the measured values of the coupling constants and the
 particle content of the standard model, one can extrapolate to higher
 energies and datermine the coupling constants there. The result is
 that the three coupling constants do not meet at the same point.
 However, repeating this extrapolation with the particles belonging to the minimal supersymmetric extension of the standard model, the three gauge coupling constants meet at a point as sketched in fig.7. 
\begin{figure}
\begin{center}
\leavevmode
\psfig{file=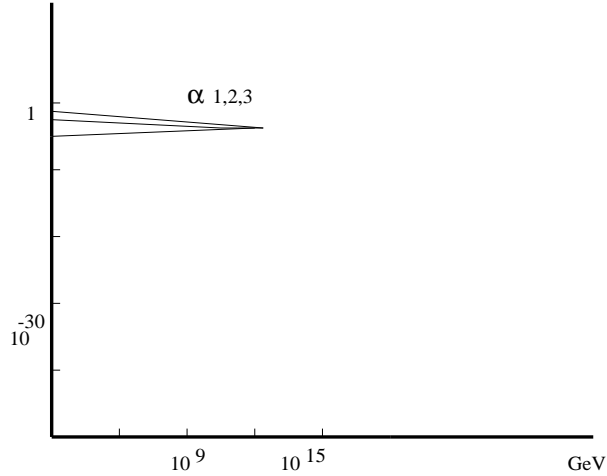,width=8cm}
\caption{The three gauge couplings as functions of the energy.}
\end{center}
\label{fip4}
\end{figure}

\subsection{Supersymmetry}
Supersymetry is a symmetry that relates bosons to fermions, though every 
fermion has a bosonic superpartner and vice versa. for exemple, fermionic 
quarks are partners of bosonic squarks. By this we mean that quarks and 
squarks belong to the same irreductible repesentation of the 
supersymmetry, if supersymmetry were an broken symmetry, particles and 
their superpartners would have exactly the same masse. So it is inherently 
quantum mechanical symmetry, since the very concept of fermions is a 
quantum mechanical.

In quantum filed theory boson fields have dimension one and fermion fields 
have $\frac{3}{2}$ in ordre that the action be dimensioneless (in units 
$\hbar = c= 1$). The reason is that boson fields have two derivatives in their 
action while fermion fields have only one. It is not difficult to see that 
two supersymmetry transformations will certainly lead to a gap of one unit 
of dimension. The only dimensional object different from fields themselves 
available to fill this gap is the derivative. Thus in any global 
supersymmetry model we can always find a derivative appearing in a double transformation relation, purely on dimensional grounds.\\
Mathematically therefore the global supersymmetry resemble taking the square root of the transformation operator. So actually it is not an internal symmetry but an enlargement of the Poincar{\'e} groupe. This amounts to an extension of space-time to {\it{superspace}} that includes extra spinorial anticommuting coordinates as well as ordinary coordinates. We do not change the structure of space-time but we add structure to it. We start with the usual coordinates $X^\mu  =t,x,y,z$ and add an odd dimensions $\theta^{\alpha} (\alpha= 1,..., n)$. These dimensions are fermionic and anticommute
\begin{equation}
\theta^{\alpha}\theta^{\beta}= - \theta^{\beta}\theta^{\alpha}
\end{equation}
They are quantum dimension that have no classical analog, which makes it difficult to visualize or to undrstand them intuitively.\\

 More formally, a genral Grassmann algebra with $n$ generators ${\cal{G}}_{n}$ is defined as follows:
\begin{itemize}
\item[i)]{${\cal{G}}_{n}$ is a vector space over the complex numbers}
\item[ii)]{a product is defined over ${\cal{G}}_{n}$, which is associative and bilinear with respect to addition and multiplication by scalars}
\item[iii)]{${\cal{G}}_{n}$ contains the unit element for this product}
\item[iv)]{${\cal{G}}_{n}$ is generated by $n$ elements $\xi^{A}, A= 1,..., n$ which obey the relation
\begin{equation}
\xi^{A}\xi^{B}+ \xi^{B}\xi^{A}= 0
\end{equation}
Furthemore, there is no other independent relations among the generators.}
\end{itemize}
It follows from the anticommutaiton of the $\xi$'s that ${\cal{G}}_{n}$ is $2^{n}$-dimensional as a vector space. A basis of ${\cal{G}}_{n}$ is given by the monomials $1, \xi^{A}, \xi^{A}\xi^{B}(A<B),...\xi^{1}\xi^{2}...\xi^{n}$. A general element $g$ of ${\cal{G}}_{n}$ reads 
\begin{equation}
g= g_{0} + g_{A}\xi^{A}+ g_{AB}\xi^{A}\xi^{B}+ ...+ g_{A_{1}A_{2}...A_{n}}\xi^{A_{1}}...\xi^{A_{n}}
\end{equation}
where the cofficients $g_{AB},..., g_{A_{1}A_{2}...A_{n}}$ can be assumed to be completely antisymmetric. The coefficients $g_{0}$ is the component of $g$ along unity.\\
So we can distinguish between even dynamical $X^\mu $ variables (commuting c-numbers) and odd dynamical variables $\theta^{\alpha}$ (anticommuting c-numbers). where
\begin{eqnarray}
X^\mu (t)&= &X_{0}^\mu (t)+ X_{AB}^{i}(t)\xi^{B}\xi^{A}+... \\
\theta^{\alpha}(t)&=& \theta^{\alpha}_{A}(t)\xi^{A}+ \theta^{\alpha}_{ABC}(t)\xi^{A}\xi^{B}\xi^{C}+...
\end{eqnarray}

A Grassman-valued function of the dynamical variables ($X^\mu , \theta^{\alpha}$) is an element of the Grassmann algebra, to which $X^\mu $ and $\theta^{\alpha}$ belong, which depends on $X^\mu $ and $\theta^{\alpha}$. In termes of components, a function ${\cal{F}}(X^\mu , \theta^{\alpha})$ is equivalent to a set of function $F_{0}, F_{A}, F_{AB}, ...$ of the components $X^\mu _{0},X^\mu _{AB}, ..., \theta^{\alpha}_{A},..$ of $X^\mu $ and $\theta^{\alpha}$ such that 
\begin{equation}
{\cal{F}}= F_{0}(X^\mu _{0},X^\mu _{AB}, ..., \theta^{\alpha}_{A},..) + F_{A}(X^\mu _{0},X^\mu _{AB}, ..., \theta^{\alpha}_{A},..)\xi^{A}+...
\end{equation}

of particular importance are the so-called {\it{superfunctions}}. These depends on the individual components $X^\mu _{0},X^\mu _{AB}, ..., \theta^{\alpha}_{A},..$ only through $X^\mu $ and $\theta^{\alpha}$ and have no explicit dependence on $\xi^{A}$.\\ \\
The fact that the odd directions are anticommuting has important consequences. Consider a function of superspace (with $\alpha = 4$)
\begin{equation}
{\cal{F}}(X, \theta)= \phi(X) + \theta_{\alpha}\psi_{\alpha}(X) + ...+ \theta^{4}F(X)
\end{equation}
Since the square of any $\theta$ is zero and there are only four different $\theta$'s the expansion in powers of $\theta$ terminates at the fourth order. Therefore, a function of superspace includes only a finite numbers of functions of $X$ (16 in this case). Hence, we can replace any function of superspace ${\cal{F}}(X,\theta)$ with the component functions $\phi(X)$, $\psi(X)$.... This include bosons $\phi(X)$ and fermions $\psi(X)$, ... . This is one way of understanding the pairing between bosons and fermions.\\ \\
A supersymmetric theory looks like an ordinary theory with degrees of freedom and interactions that satisfy certain symmetry requirement. In this sens, supersymmetry by itself is not a very radical proposal. However, the fact that bosons and fermions come in pairs in supersymmetric theories had important consequences. In some loop diagrammes the bosons and fermions cancel each other. This boson-fermion cancellation is at the heart of most of the applications of supersymmetry.\\ \\
Just as for usual symmetries, one can distinguish between two kinds of supersymmetries: global ones (rigid supersymmetry) and local ones (gauge supersymmetry). In local theory the translation operator differs from point to point. this is precisely the notion of a general coordinate transformation and leads us to expect that gravity must be present. Indeed, guided by the requirement of local supersymmetry invariance and using 'Noether's method', we can actually get massless spin $\frac{3}{2}$ field gauging supersymmetry, i.e. gravitino and massless spin-$2$ field gauging space-time symmetry, i.e. the graviton. So the local gauge theory of supersymmetry implies a local gauge theory of gravity. This is the reason for such local supersymmetry  being called supergravity. Still that in supergravity and extended supergravity the relation between supersymmetry and extra internal symmetry is still unclear. So the relation between external and internal geometries are still vague within the context of supergravity.\\ \\

On the other hand in the revitalized Kaluza-Klein theory, where the number of extra space dimensions become seven (taking into account the number of symmetry operations embodied in grand unified theories and extended N=8 supergravity), the internal symmetries are the manifestations of the geometrical symmetries associated with the extra compactified space dimensions and that all the kinds of geometries associated with internal symmetries are genuine space geometries, i.e. geometries associated with extra space dimensions. So the question concerning the relation between internal and external geometries, in a deeper sens, remains profound. Nevertheless, the modern Kaluza-Klein theory does open a door for establishing the correlation between non-gravitational gauge potentials and the geometrical srtuctures in four dimensional space-time via the geometrical structures in extra dimensions. Within this theoretical context the unifications of gravity and other gauge interactions, is in principle testable, and cannot be accused of being irrelevant to the future development of fundamental physics.\\
 In superstring the introduction of extra compactified space dimensions is due to different considerations from just reproducing the gauge symmetry. Therefore, the properties and structures of the compactified dimensions are totally different from those in the Kaluza-Klein version. For example there is no symmetry in the compact dimensions from which the gauge symmetries emerge. The gauge interactions are correlated with the geometrical structure of  ten-dimensional space-time as a whole and not with the extra dimensions. More, in ten-dimensional quantum superstring theories there are gravitational and Yang-Mills anomalies, that is the violation of the conservation of the Yang-Mills charges and the energy-momentum. Requiring the absence of all anomalies leads to requiring a very intimate relationship between gravitational and Yang-Mills interactions.
\subsection{Supersymmetry and strong coupling}

 Supersymmetry gives new information about strong coupling. To see this (we follow Polchinski \cite{polch1}), let us consider in quantum theory the Hamiltonian operator
$H$, the simplest example is the hamiltonian:
\begin{equation}
H= w_{a} a^{+}a + w_{b}b^{+}b
\end{equation}
 where we have bosonic and fermionic harmonic operators that obey:
\begin{equation}
[a,a^{+}]= [b,b^{+}]= 1
\end{equation}

The supersymmetric operator $Q$ is defined as :
\begin{equation}
Q\equiv b^{+}a + a^{+}b
\end{equation}
if $a^{+}\mid 0\rangle$ is a one boson state, then $Qa^{+}\mid 0\rangle$ becomes a one fermion state, and vice versa $Q$ obeys the following identity
\begin{equation}
[Q,H]= (w_{a} - w_{b})Q
\end{equation}
If $w_{a}=w_{b}=w$, then the supersymmetric operator $Q$ commutes with the Hamiltonian and:
\begin{equation}
\{Q,Q^{+}\}=\frac{2}{w} H
\end{equation}
These identities show that $Q$ and $Q^{+}$ form a closed algebra with the hamiltonian if the fermions and bosons have equal energy.

take in addition  the charge operator $G$\footnote{There are usually several $G$s and several $Q$s, so that
there should be additional indices and constants in these equations.} associated with an ordinary symmetry like
electric charge or baryon number.  The fact that $G$ is a symmetry means that it
commutes with the Hamiltonian,
\begin{equation}
[H, G] = 0\ .
\end{equation}
as we have said for supersymmetry one has the same,
\begin{equation}
[H, Q] = 0\ , \label{susy1}
\end{equation}
but there is an additional relation
\begin{equation}
Q^2 = H + G\ , \label{susy2}
\end{equation}
in which the Hamiltonian and ordinary symmetries appear on the right.  It is this equation that gives the extra information. 
Consider now a state $|\psi\rangle$ which is
neutral under supersymmetry:
\begin{equation}
Q |\psi\rangle = 0\ .
\end{equation}
 We
are interested in states that are neutral under {\it at least one} $Q$
but usually not all of them.  These are known as {\it BPS
(Bogomolnyi--Prasad--Sommerfield) states}. The expectation
value of the relation~(\ref{susy2}) in this state gives us:  
\begin{equation}
\langle \psi | Q^2 |\psi\rangle = \langle \psi | H |\psi\rangle + \langle
\psi | G |\psi\rangle \ .
\end{equation}
The left side vanishes by the BPS property, while the two terms on the
right are the energy $E$ of the state $|\psi\rangle$ and its charge $q$
under the operator $G$.  Thus
\begin{equation}
E = - q\ , \label{bpsen}
\end{equation}
and so the energy of the state is determined in terms of its charge. So a dynamical quantity  is determined
entirely by symmetry information.

Since the calculation of $E$ uses only symmetry information, it does not
depend on any coupling being weak. 
Thus we know something about the spectrum at strong coupling\cite{polch1}.  
The BPS states are only a small part of the spectrum, but by using this
and similar types of information from supersymmetry, together with
general properties of quantum systems, one can usually recognize a
distinctive pattern in the strongly coupled theory and so deduce the dual
theory. 

\subsection{Vacuum Selection}
Another property of many supersymmetric theories that make them tractable is that they have a family of inequivalent vacua. To understand this fact we should contrast it with the situation in ferromagnet, which has a continuum of vacua, labeled by the common orientation of the spins. These vacua are all equivalent, i.e. the physical observables in one of these vacua are exactly the same as in any other. The reason is that these vacua are related by a symmetry. The system must choose one of them, which leads to spontaneous symmetry breaking. However , in a supersymmetric theory the zero-point energy of the fermions exactly cancels that of the bosons this  is the source of the presence of some degeneracy. So we see that a supersymmetric system has a continuous family of vacua. This family, or manifold, is referred to as a moduli space of vacua.\\
The analysis of supersymmetric theories is usually simplified by the presence of these manifolds of vacua, by using the  asymptotic behavior along several directions, where the analysis of the system is simple and various approximation techniques are applicable, as well as the constraints from holomorphy\footnote{The main point is that the supersymmetric quantum field theories are very constrained. The dependance of some observables on the parameters of the problem is so constrained that there is only one solution that satisfies all the consistency conditions. More technically, because of supersymmetry some observables vary holomorphically with the coupling constants, which are complex numbers in these theories. due to Cauchy's theorem, such analytic functions are determined in terms of very little data: the singularities and the asymptotic bahavior. therefore, if supersymmetry requires an observables to depend holomorphically on the parameters and we know the singularities and the asymptotic behavior, we can determine the exact answer}, a unique solution is obtained. So an approximate calculations, which are valid only in some regime, gives us the exact answer.
 
\section{Dualities}
Few words have been used with more different meanings than the
word {\it{duality}}. Even within the restricted framework of string
theories, duality originally meant a symmetry between the s and
the t-channels in strong interactions (coming from the demands in
the S-matrix approach of the sixties of Regge behavior without fixed
poles and analiticity, which were shown to imply the existence of an
infinite number of resonances) \cite{scherk}. Somewhat related ideas,
also termed {\it{duality}}, appear in the context of Conformal Field
Theory (CFT) as simple consequences of locality and associativity
of the operator product expansion (OPE) \cite{luger}.

Duality symmetry plays an important role in
Statistical Mechanics
, in particular in
the
analysis of the phase diagram of spin systems.  It can also be
understood as a way to show the equivalence between two
apparently different theories. On a
lattice system described by a Hamiltonian $H(g_i)$
with coupling constants $g_i$
the duality transformation
produces a new Hamiltonian $H^*(g^*_i)$ with coupling
constants $g^*_i$ on the dual lattice.  In
this way one can often relate the strong coupling regime
of $H(g)$ with the weak coupling regime of $H^*(g^*)$.
An important application was the determination
of the exact temperature at which the phase transition
of the two-dimensional Ising model takes place \cite{ising}.

More recently, the word {\it{duality}} ({\it{space-time duality}}) has
been introduced in yet another sense. 
\subsection{String Duality, M-theory}
Existence of duality symmetries in string theory started out as a
conjecture and still remains a conjecture. However so many
non-trivial tests of these conjectures have been performed by now
that most people in the field are convinced of the validity of
these conjectures. \\ \\
A duality conjecture is a statement of equivalence between two or
more apparently different string theories. Two of the most
important features of duality are as follows:
\begin{itemize}
\item[i)] Often
under the duality map, an elementary particle in one theory
gets mapped to a composite particle in a dual theory and vice
versa. Thus classification of particles into elementary and
composite loses significance as it depends on which particular
theory we use to describe the system.
\item[ii)] Often duality relates a weakly coupled string theory to a
strongly coupled string theory and vice versa. In many simple
cases the coupling constants $g$ and $\wt g$ in the two theories 
are related via the simple relation:
\be \label{e1}
g = \wt g^{-1}\, .
\ee
Thus a perturbation expansion in $g$ contains information about
non-perturbative effects in the dual theory. In particular the
tree level (classical) results in one theory can contain
contribution from perturbative and non-perturbative terms in the
dual theory. This also clearly shows that duality is a property
of the full quantum string theory, and not of its classical
limit.
\end{itemize}
Thus there are two aspects of duality
\begin{center}
elementary $\leftrightarrow$ composite

classical $\leftrightarrow$ quantum
\end{center}

Let me now give some examples of dual pairs of string theories.
\begin{itemize}
\item[i)] (9+1) dimensional
SO(32) heterotic and type I string theories are conjectured
to be dual to each other.\cite{WITTEND,POLWIT}
\item[ii)] SO(32) heterotic string theory compactified on a four
dimensional torus (denoted as $T^4$)
is conjectured to be dual to type IIA
string theory compactified on a different four dimensional
manifold, denoted by $K3$.\cite{HT}
\item[iii)] Type IIB string theory is conjectured to be self-dual, in
the sense that the type IIB string theories at two different
couplings $g$ and $\wt g$ related by eq.\refb{e1} are conjectured
to describe the same physical theory.\cite{HT}
\item[iv)] Heterotic string theory compactified on a six dimensional
torus, denoted by $T^6$, is conjectured to be self-dual in the same
sense as above.\cite{FONT,SREV}
\end{itemize}

So due to the fact that a duality conjecture relates two 
apparently different theories, we see that it gives a unified
picture of all string theories. The situation is summarized in
Fig.8\footnote{
One should keep in mind that this is only a
schematic representation.}. 
\begin{figure}
\begin{center}
\leavevmode
\psfig{file=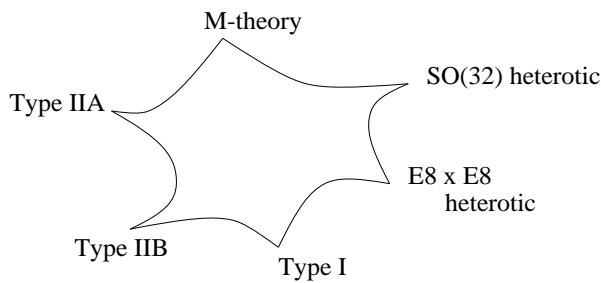,width=8cm}
\caption{The five string theories, and M-theory, as limits of a single
theory.}
\end{center}
\label{fip7}
\end{figure}

 According to this picture the apparently
different string theories and their compactifications
are just different limits of the same theory, with a large
parameter space.\footnote{In string theory parameters themselves
are related to vacuum expectation values of different fields and
are expected to be determined dynamically.} There is no
universally accepted name for this central theory, $-$ We call it $U$-theory. 
$U$ can be
taken to stand for Unknown or Unified. Some small
regions of the parameter space of
$U$-theory, which can be represented by some {\it weakly
coupled} string theory, are reasonably well understood
and correspond to the weak
coupling regime of the five different string theories and their
compactifications. But for
most of the parameter space $U$-theory does not have a description
in terms of weakly coupled string theory. Note that in one corner
of the parameter space of $U$-theory, there
is a theory called $M$-theory~\cite{TOWN,WITTEND,SCHM,ASPINM,HORWIT} 
which has not been introduced before.
At present not much is known about $M$-theory except that its low
energy limit is the eleven dimensional supergravity theory, 
and that various
string theories and their compactifications
approach $M$-theory in certain limits. However, unlike
string theory, $M$-theory does not have any coupling constant,
and no systematic procedure for doing computations in 
$M$-theory beyond the low energy supergravity limit is known. \\
 The M-theory point in the figure is in fact a point
of $SO(10,1)$ symmetry: the spacetime symmetry of string theory is larger
than had been suspected.  The extra piece is badly spontaneously broken,
at weak coupling, and not visible in the perturbation theory, but it is a
property of the exact theory.  It is interesting that $SO(10,1)$ is
known to be the largest spacetime symmetry compatible with supersymmetry.

Another way to describe this is that in the M-theory limit the theory
lives in eleven spacetime dimensions: a new dimension has appeared.  This
is one of the surprising discoveries of the past few years.

\subsection{The canonical approach to T-duality}
In String Theory and Two-Dimensional Conformal Field Theory
duality is an important tool to show the equivalence of
different geometries and/or topologies and in determining
some of the genuinely stringy implications on the structure
of the low energy Quantum Field Theory limit.
T-Duality symmetry was first described on the
context of toroidal compactifications \cite{bgs}.
For the simplest case of a single
compactified dimension of radius $R$, the entire physics
of the interacting theory is left unchanged under the
replacement $R \rightarrow {\alpha}^{'} /R $
provided one also transforms the dilaton field $\phi
\rightarrow \phi - \log{(R/\sqrt{{\alpha}^{'}})}$
\cite{ema}. This simple case can
be generalized to  arbitrary toroidal compactifications
described by constant metric $g_{ij}$ and antisymmetric
tensor $b_{ij}$ \cite{nsw}. The
generalization of duality to this case becomes
$(g+b) \rightarrow (g+b)^{-1}$ and $\phi \rightarrow \phi
-{1\over 2}\log{\mbox{det}(g+b)}$. In fact this transformation
is an element of an infinite order discrete symmetry group
$O(d,d; Z)$ for $d$-dimensional toroidal compactifications
\cite{torodd,venezia1}. The
symmetry was later extended to the case
of non-flat conformal backgrounds
in \cite{buscher}.
In Buscher's construction one starts
with a  manifold $M$ with metric
$g_{ij}, i,j=0,\ldots d-1$, antisymmetric
tensor $b_{ij}$ and  dilaton field $\phi(x_i)$.
One requires the metric to admit at least one
continuous abelian isometry leaving invariant the
$\sigma$-model action constructed out of $(g, b, \phi)$.
Choosing an adapted coordinate system $(x^0, x^{\alpha}) =
(\theta, x^{\alpha}), \alpha = 1, \ldots d-1 $
where the isometry acts by translations of $\theta$, the
change of $g, b, \phi$ is given by
\bea
\label{busdual}
&&{\tilde g}_{00}=1/g_{00},\qquad
         {\tilde g}_{0\alpha}=b_{0\alpha}/g_{00},\nonumber\\
&&{\tilde g}_{\alpha\beta}=g_{\alpha\beta} -
(g_{0\alpha}g_{0\beta} - b_{0\alpha} b_{0\beta})/g_{00}\nonumber\\
&&{\tilde b}_{0\alpha} =g_{0\alpha}/g_{00},\nonumber\\
&&{\tilde b}_{\alpha\beta}=b_{\alpha\beta}-(g_{0\alpha}b_{0\beta}
         -g_{0\beta}b_{0\alpha})/g_{00},\nonumber\\
&&\tilde {\phi}=\phi-\frac12\log{g_{00}}.
\eea

The final outcome is that for any continuous isometry of the metric
which is a symmetry of the action one obtains the equivalence
of two apparently very different non-linear $\sigma$-models.
The transformation (\ref{busdual}) is referred to in the literature
as abelian T-duality due to the abelian character of
the isometry of the original $\sigma$-model.
If $n$ is the maximal number of commuting isometries, one gets
a duality group of the form $O(n,n;Z)$
\cite{givroc}.
T-Duality symmetries are useful in determining important
properties of the low-energy effective action, in particular in
questions related to supersymmetry breaking and to the lifting of
flat directions from the potential \cite{susydual}. Although the
transformation (\ref{busdual}) was originally obtained using
a method apparently
not compatible with
general covariance, it is not difficult to modify the
construction to eliminate this drawback \cite{aagbl}.

Of more recent history is the notion of non-abelian
T-duality \cite{quevedo,givnoab,r7,aagl}, which has no analogue in
Statistical Mechanics. The basic idea of \cite{quevedo},
inspired in the treatment of abelian T-duality
presented in \cite{rocver}, is to consider a conformal
field theory with a non-abelian symmetry group $G$.\\ 

In the abelian case it is also possible to work out
the mapping between some operators in the original and
dual theories, as well as the global topology of
the dual manifold \cite{aagbl}.
Thus for $G$
abelian we have a rather thorough understanding of
the detailed local and global properties of T-duality.
In the non-abelian case global information can only be
extracted for $\sigma$-models with chiral currents \cite{aagl}.
\section{The Canonical Approach}
Some suggestions have been made in the literature
pointing (at least in the simplified situation where
all backgrounds are constant or dependent only on time)
towards an understanding
of T-duality as particular instances of canonical
transformations \cite{venezia1,venezia2}.

Following \cite{aagl2} we can show that this idea
works well when the background admits
an abelian isometry ,
laying T-duality on a simpler
setting , namely as a (privileged) subgroup of
the whole group of (non-anomalous, that is implementable
in Quantum Field Theory \cite{ghandour})
canonical transformations on the phase space of the theory.

So Buscher's transformation formulae can be
derived by
performing a given canonical transformation on the Hamiltonian
of the
initial theory. This is a {\it{minimal}}
approach in
the sense that no extraneous structure has to be introduced,
and all
standard results in the abelian case  are easily
recovered
using it. In particular it is possible to perform the
T-duality
transformation in arbitrary coordinates not only in the
original
manifold (which was also possible in Ro\u{c}ek and
Verlinde's
formulation) but also in the dual one. Even more,
all the generators
of the full T-duality group $O(d,d;Z)$ can be described
in terms of canonical transformations.
This gives the impression that the T-duality group
should be understood in terms of global symplectic
diffemorphisms. It would be useful to formulate
it in the context of some analogue of the
group of disconnected diffeomorphisms, but for the
time being such a construction is lacking.

Concerning non-abelian duality, it seems to
fall beyond the scope of the Hamiltonian point
of view. There is one example \cite{zachos} in which
the non-abelian
dual has been constructed out of a canonical
transformation
but it is still early to say whether the general case can be
treated similarly.

\subsection{The Abelian Case}
\setcounter{equation}{0}

We start with a bosonic
sigma model written in arbitrary coordinates
on a manifold $M$ with Lagrangian
\be
\label{sigmamodel}
L = \frac12 (g_{ab}+b_{ab})(\phi) \partial_{+}\phi^a
\partial_{-}\phi^b
\ee
where $x^\pm =(\tau\pm \sigma)/2$, $a,b=1,\dots ,d={\rm dim}M$.
The corresponding Hamiltonian is
\be
H=\frac12 (g^{ab} (p_a-b_{ac}\phi^{'}\,^c)
(p_b-b_{bd}\phi^{'}\,^d) + g_{ab}\phi^{'}\,^a
\phi^{'}\,^b)
\ee
where $\phi^{'}\,^a\equiv d\phi^a/d\sigma$.
We assume moreover that there is a Killing vector
field $k^a$,
${\cal L}_k g_{ab}=0$ and $i_k H=-dv$ for some
one-form $v$,
where $(i_k H)_{ab}\equiv k^c H_{cab}$
and $H=db$ locally.
This guarantees the existence of a particular
system of coordinates, ``adapted coordinates'',
which we
denote by $x^i\equiv (\theta,x^\alpha)$, such that
$\vec{k}=\partial / \partial\theta$. We denote the
jacobian matrix by $e^i_a\equiv\partial x^i/
\partial \phi^a$.

This defines a point transformation
in the original Lagrangian (\ref{sigmamodel})
which acts on the Hamiltonian as a canonical
transformation
with generating
function $\Phi=x^i(\phi)p_i$, and yields:
\bea
\label{candual1}
&&p_a=e^i_a p_i \nonumber\\
&&x^i=x^i(\phi).
\eea
Once in adapted coordinates we can write the sigma model
Lagrangian as
\be
L=\frac12 G (\dot{\theta}^2-\theta^{'}\,^2)+(\dot{\theta}
+\theta^{'})J_-+
(\dot{\theta}-\theta^{'})J_++V
\ee
where
\bea
G=g_{00}=k^2 \qquad V=\frac12 (g_{\alpha\beta}+
b_{\alpha\beta})\partial_+
x^\alpha\partial_-x^\beta
\nonumber\\
J_-=\frac12 (g_{0\alpha}+b_{0\alpha})\partial_-x^\alpha
\qquad J_+=\frac12
(g_{0\alpha}-b_{0\alpha})\partial_+x^\alpha.
\eea
In finding the dual with a canonical
transformation we can use the Routh function with respect
to $\theta$,
i.e. we only apply the Legendre transformation to
$(\theta,\dot{\theta})$. The canonical momentum is given by
\be
p_\theta=G\dot{\theta}+(J_++J_-)
\ee
and the Hamiltonian
\bea
\label{hamil}
&&H=p_\theta \dot{\theta}-L=\frac12 G^{-1} p_\theta^2-
G^{-1}(J_++J_-)p_\theta+\frac12 G\theta^{'}\,^2+ \nonumber\\
&&+\frac12 G^{-1}(J_++J_-)^2+\theta^{'}(J_+-J_-)-V.
\eea
The Hamilton equations are:
\bea
\dot{\theta}=\frac{\delta H}{\delta p_\theta}=
G^{-1}(p_\theta-J_+-J_-) \nonumber\\
\dot{p_\theta}=-\frac{\delta H}{\delta\theta}=
(G\theta^{'}+J_+-J_-)^{'}
\eea

The generator of the
canonical transformation we choose is:
\be
\label{fungen}
F=\frac12 \int_{D, \partial D=S^1} d\tilde{\theta}
\wedge d\theta=
\frac12 \oint_{S^1}
(\theta^{'}\tilde{\theta}-\theta\tilde{\theta}^{'})
d\sigma
\ee
that is,
\bea
\label{fungen2}
&&p_\theta=\frac{\delta F}{\delta\theta}=-
\tilde{\theta}^{'} \nonumber\\
&&p_{\tilde{\theta}}=-\frac{\delta F}{\delta\tilde{\theta}}
=-\theta^{'}.
\eea
This generating functional does not receive any quantum
corrections
(as explained in \cite{ghandour}) since it is linear
in $\theta$ and
$\tilde{\theta}$.  If $\theta$ was not an adapted
coordinate to
a continuous isometry, the canonical transformation would
generically lead to a non-local form of the dual Hamiltonian.
Since the Lagrangian and Hamiltonian in our case only depend
on the time- and space-derivatives of $\theta$, there are no
problems with non-locality. The transformation (\ref{fungen2})
in (\ref{hamil}) gives:
\bea
\label{canoni}
&&\tilde{H}=\frac12 G^{-1} \tilde{\theta}^{'}\,^2+
G^{-1}(J_++J_-)
\tilde{\theta}^{'}+ \nonumber\\
&&\frac12 G p_{\tilde{\theta}}^2-(J_+-J_-)
p_{\tilde{\theta}}+
\frac12 G^{-1} (J_++J_-)^2-V.
\eea
Since:
\be
\dot{\tilde{\theta}}=\frac{\delta\tilde{H}}{\delta
p_{\tilde{\theta}}}=Gp_{\tilde{\theta}}-(J_+-J_-),
\ee
we can perform the inverse Legendre transform:
\bea
&&\tilde{L}=\frac12 G^{-1} (\dot{\tilde{\theta}}^2-
\tilde{\theta}^{'}\,^2)+
G^{-1}J_+(\dot{\tilde{\theta}}-\tilde{\theta}^{'}) \nonumber\\
&&-G^{-1}J_-(\dot{\tilde{\theta}}+\tilde{\theta}^{'})+
V-2G^{-1}J_+J_-.
\eea
{}From this expression we can read the dual metric and torsion
and check that they are given by Buscher's
formulae\footnote{The
minus
signs in $\tilde{g}_{0\alpha}$ and
$\tilde{b}_{0\alpha}$
can be absorbed
in a redefinition $\tilde{\theta}\rightarrow -
\tilde{\theta}$.}:
\bea
\label{buscher}
&&\tilde{g}_{00}=1/g_{00},\qquad
         \tilde{g}_{0\alpha}=-b_{0\alpha}/g_{00},\qquad
          \tilde{g}_{\alpha\beta} = g_{\alpha\beta} -
\frac{g_{0\alpha}g_{0\beta} - b_{0\alpha} b_{0\beta}}{g_{00}}
\nonumber\\
&&\tilde{b}_{0\alpha} = -\frac{g_{0\alpha}}{g_{00}},\qquad
        \tilde{b}_{\alpha\beta}=b_{\alpha\beta}-
\frac{g_{0\alpha}b_{0\beta}-g_{0\beta}b_{0\alpha}}{g_{00}}
\eea
For the dual theory to be conformal invariant the dilaton
must transform as $\Phi^{'}=\Phi-\frac12\log{g_{00}}$
\cite{buscher} \cite{ema}.\\ \\

The dual manifold $\tilde{M}$ is automatically
expressed in coordinates adapted to the dual Killing vector
$\tilde{\vec{k}}=\partial / \partial\tilde{\theta}$.
We can now
perform
another point transformation, with the same jacobian as
(\ref{candual1})
to express the dual manifold in coordinates which are as
close as
possible to the original ones.

The transformations we perform are then: First a point
transformation $\phi^a\rightarrow \{\theta, x^\alpha\}$,
to go
to adapted coordinates in the original manifold. Then a
canonical transformation $\{\theta, x^\alpha\}\rightarrow
\{\tilde{\theta}, x^\alpha\}$,
which is the true duality transformation. And finally
another
point transformation $\{\tilde{\theta}, x^\alpha\}
\rightarrow
\tilde{\phi}^a$,
with the same jacobian as the first point transformation,
to express
the dual manifold in general coordinates.

It turns out that the composition of these three
transformations
can be expressed in geometrical terms using only the
Killing
vector
$k^a$, $\omega_a\equiv e^0_a$ and the corresponding dual
quantities\footnote{Note that we must raise and
lower indices
with the dual metric, i.e. $\tilde{e}_{ia}=\tilde{g}_{ij}
\tilde{e}^{j}_a, \tilde{e}^{ia}=\tilde{g}^{ab}
\tilde{e}^{i}_b$,
which implies $\tilde{\omega}_a=\omega_a$, but
$\tilde{\omega}^a=k^a (k^2+v^2)+\vec{e}\,^a
\cdot v$ (where
$\vec{e}\,^a\equiv e^a_{\alpha}$),
$\tilde{k}^a=k^a$ but
$\tilde{k}_a=(\omega_a -
(\vec{e}_a\cdot v))/k^2$. We have
moreover $\tilde{\omega}\,^2=
k^2+v^2+g^{\alpha\beta} v_{\beta}{\omega}_{\alpha}$ and
$\tilde{k}^2=1/k^2$.}.

The canonical approach has been very useful in order
to obtain the dual manifold in an arbitrary coordinate system.
With the usual approaches it is expressed in
adapted coordinates to the dual isometry. This happens
because the
dual variables appear as Lagrange multipliers and after an
integration by parts only the derivatives of them emerge,
being then adapted coordinates automatically.

\subsection{Gauge Theories from Branes}

The study of branes has been useful in
deriving gauge theory results from string
theory\cite{FTH,BADOSE,HANWIT,WITBRANE}.

A $p$-brane denotes a static
configuration which
extends along $p$ spatial direction (the tangential directions) and 
is localized in all other
spatial directions (the transverse directions). 
Thus the solution is invariant under
translation along the $p$ directions tangential to the brane, as
well as the time direction, and
approaches the vacuum configuration as we go away from the brane
in any one of the transverse direction. 
Thus in this language, 
\begin{eqnarray}
0-brane \qquad &\equiv& \qquad particle   \nonumber\\
1-brane \qquad &\equiv& \qquad string   \nonumber  \\
2-brane \qquad &\equiv& \qquad membrane \nonumber
\end{eqnarray}
etc.
Typically the quantum
dynamics of a configuration of $p$-branes is described by a
$(p+1)$ dimensional gauge field theory,\cite{POLDBR,WITDBR} and 
the coupling
constant of this quantum field theory is related to the coupling
constant of the string theory of which the brane configuration is
a solution. In this case duality symmetries relating strong and
weak coupling limits of the original string theory can be used to
derive duality relations involving the quantum field theories
describing the dynamics of the brane. This approach has been used
to derive many different results in supersymmetric gauge
theories. Some example are:
\begin{itemize}
\item[i)] Derivation of Montonen-Olive duality~\cite{MONOL} in $\NN=4$
supersymmetric gauge theories.\cite{NEFOUR}
\item[ii)] Derivation of Seiberg-Witten like results~\cite{SEIWIT} 
in $\NN=2$
supersymmetric gauge theories.\cite{FTH,BADOSE,WITBRANE}

\item[iii)]
Derivation of a special kind of symmetry, known as the mirror
symmetry,\cite{MIRROR} 
in (2+1) dimensional gauge theories.\cite{HANWIT}

\item[iv)]
Derivation of Seiberg dualities~\cite{SEIBERG} involving 
$\NN=1$ supersymmetric
gauge theories in (3+1) dimensions.\cite{NEQONE,WITNEONE}
\end{itemize}

A special class of $p$-branes are called Dirichlet $p$-branes (D-branes)\footnote{quantum versions of objects that were first found as solitonic solutions of supergravity}. The name derives from the boundary conditions assigned to the ends of open strings. More general, in type II theories, one can consider an open string boundary conditions at the end given by $\sigma = 0$
\begin{eqnarray}
\frac{\partial X^\mu }{\partial \sigma}= 0 \,\, \mu = 0, 1, ..., p  \\
X^\mu  = X^\mu _{0} \, \, \mu = p+1, ..., 9
\end{eqnarray}

and similar boundary conditions at the other end\footnote{The interpretation of these equations is that strings end on a $p$-dimensional object in space- a $D$-branes. The description of $D$-branes as a place where open strings can end leads to a simple picture of their dynamics. For weak string coupling this enables the use of perturbation theory to study non-perturbative phenomena. One of the most remarkable of these concerns the study of black holes. D-branes have a very strange property the description of their positions suggest that the space-time coordinates must be reinterpreted as noncommuting matrices.}. fig.9.
\begin{figure}
\begin{center}
\leavevmode
\psfig{file=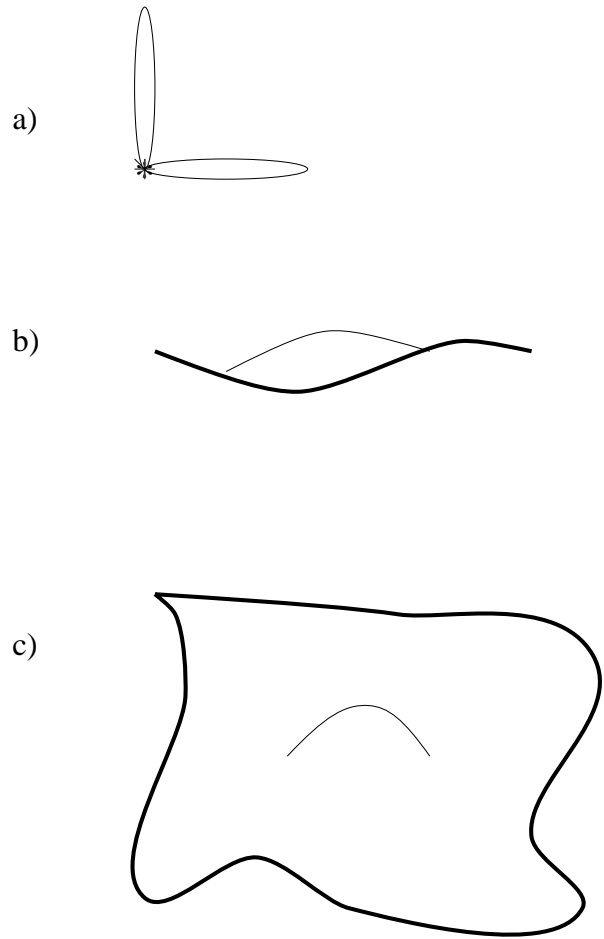,width=8cm}
\caption{a) A D0-brane with two attached strings.  b) A D1-brane (bold)
with attached string.  c) A D2-brane with attached string.}
\end{center}
\label{fip}
\end{figure}

Existence of branes in string theory has also given rise to the
possibility that the standard model gauge fields arise from
branes rather than in the bulk of space-time. This corresponds to
novel compactifications in which gravity lives in
the bulk of the ten dimensional space-time, but the other
observed fields (quarks, leptons, gauge particles etc.) live on a
brane of lower dimension\cite{SEIBWIT}. 

\subsection{Maldacena Conjecture}
A $p$-brane, or collection of $p$-branes, gives rise to a certain space-time
geometry and gauge field configuration, which can be analyzed using the
appropriate supergravity field equations.  In a number of cases one finds that
the geometry has an event horizon, giving a higher-dimensional analog
of black holes.  In some of these cases the geometry near the horizon is
approximated by $AdS_{p+2} \times S^{D-p-2}$.  This means that the AdS space
has $p + 2$ dimensions and the remainder of the $D$ dimensions form a sphere of $D -
p - 2$ dimensions.  There are three basic examples that have maximal
supersymmetry (32 conserved supercharges).  A stack of D3 branes in type IIB
superstring theory has near horizon geometry $AdS_5 \times S^5$, a stack of
M2-branes in M theory gives $AdS_4 \times S^7$, and a stack of M5-branes in M
theory gives $AdS_7 \times S^4$.  \\
Let me briefly describe some  features of anti de Sitter space.
$AdS_{n+1}$ is a maximally symmetric spacetime with a negative cosmological
constant.
It can be described as a hypersurface in flat space by the equation
\begin{equation}
u_1^2 + u_2^2 - v_1^2 - v_2^2 - \ldots - v_n^2 = R^2,
\end{equation}
where $R$ is called the AdS radius.  This spacetime has Lorentzian signature
and reduces to Minkowski spacetime in $n + 1$ dimensions in the limit $R
\rightarrow \infty$.  Just as an $n+1$-dimensional 
sphere ($S^{n+1}$) has $SO(n+2)$ symmetry, the
symmetry of this spacetime is $SO(2, n)$, a noncompact version of the rotation
group in $n + 2$ dimensions.  This contracts to the Poincar{\'e}
group (consisting of the Lorentz group $SO(1,n)$ and translations)
in the $R \rightarrow \infty$ limit.  An
intrinsic description of $AdS_{n+1}$ is given by the metric
\begin{equation} \label{metric}
ds^2 = \frac{R^2}{z^2} (dz^2 + dx^\mu dx_\mu ),\quad  z>0,
\end{equation}
where
\begin{equation}
dx^\mu dx_\mu = dx_1^2 + \ldots + dx_{n-1}^2 - dt^2.
\end{equation}
Note that the $z = 0$ boundary of $AdS_{n + 1}$ is an $n$-dimensional Minkowski
spacetime, aside from a divergent factor.  What matters is the {\em conformal}
structure, which is not sensitive to this divergent factor.  

The $SO(2,n)$ isometries of the $(n+1)$-dimensional anti de Sitter space induce
the group of conformal transformations on its $n$-dimensional Minkowski
boundary.  (Strictly speaking, the boundary should be compactified by adding a
point at infinity.)  The conformal group is therefore also $SO(2,n)$.  Let me
illustrate how this works with a couple of examples.  The $SO(1,n-1)$ subgroup
of $SO(2,n)$ given by Lorentz transformations of the $x^\mu $ corresponds to the
Lorentz group of the boundary.  The important point is that these
transformations map $z = 0$ to $z = 0$, so that they are well-defined on the
boundary.  Another example is the isometry $x^\mu \rightarrow \lambda x^\mu , z
\rightarrow \lambda z$ where $\lambda$ is a positive scale factor.  This
clearly leaves the AdS metric in eq.~(\ref{metric}) invariant and preserves the boundary.  Thus the
corresponding conformal transformations of the boundary are scale
transformations $x^\mu \rightarrow \lambda x^\mu $.

The basic idea of AdS/CFT duality is to identify a conformally invariant field
theory (CFT) on the $n$-dimensional boundary with a suitable quantum gravity theory in
the $(n + 1)$-dimensional AdS {\em bulk}.  \\
 The IIB theory contains a four-index field
$A_{\mu \nu\rho\lambda}$ for which the D3-brane is a source.  It has a field
strength $F_{\mu \nu\rho\lambda\sigma}$, which is self-dual (in ten dimensions).
 In the $AdS_5 \times S^5$ solution of the theory, the field $F$ has a
quantized flux on the sphere.  Schematically,
\begin{equation}
\int_{S^{5}} F = N,
\end{equation}
where $N$ is a positive integer.  This integer determines the radius $R$ of the
$AdS_5$ and of the $S^5$, which are the same.  Aside from a constant numerical
factor, one finds that
\begin{equation}
R = (g_s N)^{1/4} \ell_s.
\end{equation}
Thus the curvatures (which are proportional to $R^{-2}$)
are small compared to the string scale for $g_s N \gg 1$ and small
compared to the Planck scale for $N\gg  1$.  The first limit suppresses stringy
corrections to supergravity, whereas the latter suppresses quantum corrections
to classical string theory.

Maldacena's duality conjecture is that type IIB superstring theory on $AdS_5
\times S^5$ with $N$ units of $F$ flux {\em is equivalent to} ${\mathcal N} =
4, $ $D = 3 + 1 $ $U(N)$ Yang--Mills theory with $g_{YM}^2 = g_s$.  For this
conjecture to be plausible, it is a crucial fact the ${\mathcal N} = 4$ super
Yang--Mills theory~\cite{brink} is a CFT with vanishing beta function, a fact that was
proved in the early 1980s~\cite{mandelstam}.  
As should be clear from our presentation, this
conjecture arose from considering the near-horizon geometry of a stack of $N$
D3-branes, in the limit $N \rightarrow \infty$.

\section{conclusion}
The underlying conception of space-time advocated by string theories is very interesting. Yet the conservative side of its conception is also striking: it takes the enlarged space-time coordinates themeselves directly as physical degrees of freedom that have to be treated dynamically and quantum mechanically. Nevertheless, superstring theory gives an existence proof that gravity can be treated quantum mechanically in a consistent way. More, it has predicted many new physical degrees of freedom: strings, classical objects such as smooth solitons and singular black hole, and new types of topological defects such as D-branes. It sheds much light on field theory duality and on the gauge-theoretical structure of the standard model.\\
In field theory, we now understand that not just confinement but the whole range of surprise of strongly coupled field theory should be derived from duality, at least in the supersymmetric case. For string theory the change in viewpoint is perhaps even wider and includes the discovery that there is only one theory. To understand the dualities, or the relations between the different string theories, we have had to learn about this new degrees of freedom in string theory, such as D-branes .


\begin{thebibliography}{99}

\bibitem{GSW}
M. Green, J. Schwarz and E. Witten,  Superstring Theory vol. 1
and 2, Cambridge University Press (1986); \\
D. Lust and S. Theisen, Lectures on String Theory, Springer
(1989); \\
J. Polchinski, hep-th/9411028.

\bibitem{AS}
A. Sen, [hep-th/9802051].

\bibitem{WITTEND}
E. Witten, Nucl. Phys. {\bf B443} (1995) 85 [hep-th/9503124].

\bibitem{POLWIT}
J. Polchinski and E. Witten, Nucl. Phys. {\bf B460} (1996) 525
[hep-th/9510169].

\bibitem{HT}
C. Hull and P. Townsend, Nucl. Phys. {\bf B451} (1995) 525
[hep-th/9505073].

\bibitem{FONT}
A. Font, L. Ibanez, D. Lust and F. Quevedo, Phys. Lett. {\bf
B249} (1990) 35; \\
S. Rey, Phys. Rev. {\bf D43} (1991) 526.

\bibitem{SREV}
A. Sen, Int. J. Mod. Phys. {\bf A9} (1994) 3707 [hep-th/9402002]
and references therein.

\bibitem{DABHAR}
A. Dabholkar and J. Harvey, [hep-th/9809122].

\bibitem{TOWN}
P. Townsend, Phys. Lett. {\bf 350B} (1995) 184 [hep-th/9501068]. 

\bibitem{SCHM}
J. Schwarz, Phys. Lett. {\bf B367} (1996) 97 [hep-th/9510086].

\bibitem{ASPINM}
P. Aspinwall, Nucl. Phys. Proc. Suppl. {\bf 46} (1996) 30
[hep-th/9508154].

\bibitem{HORWIT}
P. Horava and E. Witten, Nucl. Phys. {\bf B460} (1996) 506
[hep-th/9510209]; Nucl. Phys. {\bf B475} (1996) 94
[hep-th/9603142].

\bibitem{ASDUAL}
A. Sen Phys. Lett. {\bf B329} (1994) 217 [hep-th/9402032].

\bibitem{VWIT}
C. Vafa and E. Witten, Nucl. Phys. {\bf B431} (1994) 3
[hep-th/9408074].

\bibitem{KAPL}
S. Kachru and C. Vafa, Nucl.Phys. {\bf B450} (1995) 69
[hep-th/9605105]; \\
V. Kaplunovsky, J. Louis and S. Theisen, Phys. Lett. {\bf B357}
(1995) 71 [hep-th/9506110].

\bibitem{HAWK}
S. Hawking, Nature {\bf 248} (1974) 30; Comm. Math. Phys. {\bf
43} (1975) 199.

\bibitem{BEKEN}
J. Bekenstein, Lett. Nuov. Cim. {\bf 4} (1972) 737; Phys. Rev.
{\bf D7} (1973) 2333; Phys. Rev. {\bf D9} (1974) 3192; \\
G. Gibbons and S. Hawking, Phys. Rev. {\bf D15} (1977) 2752..

\bibitem{SUSSBH}
L. Susskind, [hep-th/9309145]; \\
L. Susskind and J. Uglam, Phys. Rev. {\bf D50} (1994) 2700
[hep-th/9401070]; \\
J. Russo and L. Susskind, Nucl. Phys. {\bf B437} (1995) 611
[hep-th/9405117].

\bibitem{ASBH}
A. Sen, Mod. Phys. Lett. {\bf A10} (1995) 2081 [hep-th/9504147];
\\
A. Peet, Nucl. Phys. {\bf B456} (1995) 732 [hep-th/9506200].

\bibitem{STRVAF}
A. Strominger and C. Vafa, Phys. Lett. {\bf B379} (1996) 99
[hep-th/9601029].

\bibitem{CALMAL}
C. Callan and J. Maldacena, Nucl. Phys. {\bf B472} (1996) 591
[hep-th/9602043].

\bibitem{MDW}
A. Dhar, G. Mandal and S. Wadia, Phys. Lett. {\bf B388} (1996) 51
[hep-th/9605234].

\bibitem{DASMAT}
S. Das and S. Mathur, Phys. Lett. {\bf B375} (1996) 103
[hep-th/9601152].

\bibitem{STRMAL}
J. Maldacena and A. Strominger, Phys. Rev. {\bf D55} 
(1997) 861 [hep-th/9609026].

\bibitem{GREE}
M. Green and M. Gutperle, [hep-th/9701093]; \\
M. Green and P. Vanhove, Phys. Lett. {\bf B408} (1997) 122
[hep-th/9704145]; \\
M. Green, M. Gutperle and P. Vanhove, Phys. Lett. {\bf B409}
(1997) 177 [hep-th/9706175]; \\
M. Green, M.Gutperle and H. Kwon, Phys. Lett. {\bf B421} (1998)
149 [hep-th/9710151]; \\
M. Green and M. Gutperle, Phys. Rev. {\bf D58} (1998) 046007
[hep-th/9804123]; \\
M. Green and S. Sethi, [hep-th/9808061].


\bibitem{BERVAF}
N. Berkovits and C. Vafa, [hep-th/9803145].

\bibitem{KIRPIO}
E. Kiritsis and B. Pioline, Nucl. Phys. {\bf B508} (1997) 509
[hep-th/9707018]; Phys. Lett. {\bf B418} (1998) 61
[hep-th/9710078]; \\
B. Pioline, Phys. Lett. {\bf B431} (1998) 73 [hep-th/9804023]; \\
J. Russo and A. Tseytlin, Nucl.Phys. {\bf B508} (1997) 245
[hep-th/9707134]; \\
J. Russo, Phys. Lett. {\bf B417} (1998) 253 [hep-th/9707241];
[hep-th/9802090]; \\
A. Kehagias and H. Partouche, Phys. Lett. {\bf B422} (1998) 109
[hep-th/9710023].

\bibitem{BFSS}
T. Banks, W. Fischler, S. Shenker and L. Susskind, Phys. Rev.
{\bf D55} (1997) 5112 [hep-th/9610043].

\bibitem{WT}
W. Taylor, hep-th/9801182.

\bibitem{polch1}
J. Polchinski, hep-th/9812104
\bibitem{SUSS}
L. Susskind, [hep-th/9704080].

\bibitem{ASMAT}
A. Sen, Adv. Theor. Math. Phys. {\bf 2} (1998) 51 [hep-th/9709220].

\bibitem{SEIMAT}
N. Seiberg, Phys. Rev. Lett. {\bf 79} (1997) 3577 [hep-th/9710009].

\bibitem{BECBEC}
K. Becker and M. Becker, Nucl. Phys. {\bf B506} (1997) 48
[hep-th/9705091].

\bibitem{BBPT}
K. Becker, M. Becker, J. Polchinski and A. Tseytlin, Phys. Rev.
{\bf D56} (1997) 3174 [hep-th/9706072].

\bibitem{VERVER}
R. Dijkgraaf, E. Verlinde and H.Verlinde, Nucl. Phys. {\bf B500}
(1997) 43 [hep-th/9703030].

\bibitem{OKAYON}
Y. Okawa and T. Yoneya, [hep-th/9806108]; \\
W. Taylor and M. Van Raamsdonk, [hep-th/9806066]; \\
M. Fabbrichesi, G. Ferretti and R. Iengo, JHEP06(1998)2
[hep-th/9806012]; [hep-th/9806166]; \\
J. McCarthy, L. Susskind and A. Wilkins, [hep-th/9806136]; \\
U. Danielsson, M. Kruczenski and P. Stjernberg,
[hep-th/9807071]; \\
R. Echols and J. Grey, [hep-th/9806109].


\bibitem{DOS}
M. Douglas, H. Ooguri and S. Shenker, Phys. Lett. {\bf B402}
(1997) 36 [hep-th/9702203]; \\
M. Douglas and H. Ooguri, [hep-th/9710178].

\bibitem{KABPOL}
D. Kabat and W. Taylor, Phys. Lett. {\bf B426} (1998) 297
[hep-th/9712185].

\bibitem{DDM}
J. David, A. Dhar and G. Mandal, [hep-th/9707132].

\bibitem{DINE}
M. Dine, R. Echols and J. Grey, [hep-th/9810021].

\bibitem{MALD}
J. Maldacena, [hep-th/9711200].

\bibitem{OSBORN}
L. Brink, J. Schwarz and J.Scherk, Nucl. Phys. {\bf B121} (1977)
77.

\bibitem{WITONE}
E. Witten, [hep-th/9802150].

\bibitem{GUPOKL}
S. Gubser, A. Polyakov and I. Klebanov, [hep-th/9802109].

\bibitem{HOOF}
G. 't Hooft, [gr-qc/9310026].

\bibitem{SUSSHOL}
L. Susskind, J. Math. Phys. {\bf 36} (1995) 6377.

\bibitem{SUSWIT}
L. Susskind and E. Witten, [hep-th/9805114].

\bibitem{HOOFN}
G. 't Hooft, Nucl. Phys. {\bf B72} (1974) 461.

\bibitem{MALWIL}
J. Maldacena, [hep-th/9803002].

\bibitem{REYEE}
S. Rey and J. Yee, [hep-th/9803001].

\bibitem{WITTWO}
E. Witten, [hep-th/9803131].

\bibitem{NUMERIC}
C. Csaki, H. Ooguri, Y. Oz and J. Terning, [hep-th/9806021]; \\
R. de Mello Koch, A. Jevicki, M. Mihailescu and J. Nunes,
[hep-th/9806125]; \\
M. Zyskin, [hep-th/9806128].

\bibitem{FTH}
A. Sen, Nucl. Phys. {\bf B475} (1996) 562 [hep-th/9605150].

\bibitem{BADOSE}
T. Banks, M. Douglas and N. Seiberg, Phys. Lett. {\bf B387}
(1996) 278 [hep-th/9605199].

\bibitem{HANWIT}
A. Hanany and E. Witten, Nucl. Phys. {\bf B492} (1997) 152
[hep-th/9611230].

\bibitem{WITBRANE}
E. Witten, Nucl. Phys. {\bf B500} (1997) 3 [hep-th/9703166].

\bibitem{POLDBR}
J. Polchinski, Phys. Rev. Lett. {\bf 75} (1995) 4724
[hep-th/9510017].

\bibitem{WITDBR}
E. Witten, Nucl. Phys. {\bf B460} (1996) 335 [hep-th/9510135].

\bibitem{MONOL}
C. Montonen and D. Olive, Phys. Lett. {\bf B72} (1977) 117; \\
H. Osborn, Phys. Lett. {\bf B83} (1979) 321. 

\bibitem{NEFOUR}
P. Townsend, Phys. Lett. {\bf B373} (1996) 68
[hep-th/9512062]; \\
M. Green and M. Gutperle, Phys. Lett. {\bf B377} (1996) 28
[hep-th/9602077]; \\
A. Tseytlin, [hep-th/9602064].

\bibitem{SEIWIT}
N. Seiberg and E. Witten, Nucl. Phys. {\bf B426} (1994) 19
[hep-th/9407087];
Nucl. Phys. {\bf B431} (1994) 
484 [hep-th/9408099].

\bibitem{MIRROR}
K. Intrilligator and N. Seiberg, Phys. Lett. {\bf B387} (1996)
513 [hep-th/9607207].

\bibitem{SEIBERG}
N. Seiberg, Nucl. Phys. {\bf B435} (1995) 129 [hep-th/9411149].

\bibitem{NEQONE}
S. Elizur, A.Giveon and D. Kutasov, Phys. Lett. {\bf B400} (1997)
269 [hep-th/9702014]; \\
S. Elizur, A. Giveon, D. Kutasov, E. Rabinovici and A. Schwimmer,
Nucl. Phys. {\bf B505} (1997) 202 [hep-th/9704104].

\bibitem{WITNEONE}
E. Witten, Nucl. Phys. {\bf B507} (1997) 658 [hep-th/9706109].

\bibitem{SEIBWIT}
E. Witten, Nucl. Phys. {\bf B471} (1996) 135 [hep-th/9602070].

\bibitem{KAPLUN}
E. Caceres, V. Kaplunovsky, I. Mandelberg, Nucl. Phys. {\bf B493}
(1997) 73 [hep-th/9606036].

\bibitem{DIM}
N. Arkani-Hamed, S. Dimopoulos and G. Dvali, Phys. Lett. {\bf
B429} (1998) 263 [hep-ph/9803315]; \\
I. Antoniadis, 
N. Arkani-Hamed, S. Dimopoulos and G. Dvali, [hep-ph/9804398]; \\
K. Dienes, E. Dudas and T. Gherghetta, [hep-ph/9803466],
[hep-ph/9806292].






\bibitem{SEIBWIT2}
E. Witten, Nucl. Phys. {\bf B471} (1996) 135 [hep-th/9602070].

\bibitem{KAPLUN}
E. Caceres, V. Kaplunovsky, I. Mandelberg, Nucl. Phys. {\bf B493}
(1997) 73 [hep-th/9606036].

\bibitem{DIM}
N. Arkani-Hamed, S. Dimopoulos and G. Dvali, Phys. Lett. {\bf
B429} (1998) 263 [hep-ph/9803315]; \\
I. Antoniadis, 
N. Arkani-Hamed, S. Dimopoulos and G. Dvali, [hep-ph/9804398];
\bibitem{scherk}
J. Scherk, Rev. Mod. Phys. 47 (1975) 123.

\bibitem{luger}
L. {\'A}lvarez-Gaum{\'e}, G. Sierra and C. G{\'o}mez,
{\it Topics in Conformal Field Theory}, in Knizhnik
Memorial Volume,
Eds. L. Brink, D. Friedan and A.M. Polyakov,
World Scientific, Singapore, 1989.

\bibitem{dizyk}
J.M. Drouffe
and C. Itzykson, {\it Quantum Field Theory and Statistical
Mechanics},
Cambridge U. Press, 1990.

\bibitem{ising}
H.A. Kramers and G.H. Wannier, Phys. Rev. 60 (1941) 252.

\bibitem{porrati}
A. Giveon, M. Porrati and E. Rabinovici, Phys. Rep. 244
(1994) 77.

\bibitem{montonen}
C. Montonen and D. Olive, Phys. Lett. B72 (1977) 117.

\bibitem{sen}
A. Sen, Int. J. Mod. Phys. A9 (1994) 3707.

\bibitem{ot}
K.H. O'Brien and C.-I Tan, Phys. Rev. D36 (1987) 1184;
E. {\'A}lvarez and M.A.R. Osorio, Nucl. Phys. B304 (1988)
327;
J.J. Atick and E. Witten, Nucl. Phys. B310 (1988) 291.

\bibitem{bgs}
L. Brink,
M.B. Green and J.H. Schwarz, Nucl. Phys. B198 (1982) 474;
K. Kikkawa and M. Yamasaki,
Phys. Lett. B149 (1984) 357; N. Sakai
and I. Senda, Prog. Theor. Phys. 75 (1984) 692.

\bibitem{ema} E. {\'A}lvarez and M.A.R. Osorio, Phys. Rev.
D40 (1989) 1150.

\bibitem{narain}
K. Narain, Phys. Lett. B169 (1986) 41.

\bibitem{nsw} K. Narain, H. Sarmadi and E. Witten,
Nucl. Phys. B279 (1987) 369.

\bibitem{torodd}
A. Shapere and F. Wilczek, Nucl. Phys. B320 (1989) 609.

\bibitem{venezia1}
A. Giveon, E. Rabinovici and G. Veneziano, Nucl. Phys. B322
(1989) 167.

\bibitem{susydual}
A. Font,
L.E. Iba{\~n}ez, D. L{\"u}st and F. Quevedo, Phys. Lett. B245
(1990) 401; S. Ferrara, N. Magnoli,
T.R. Taylor and G. Veneziano, Phys. Lett. B245
(1990) 409; H.P. Nilles and M. Olechowski,
Phys. Lett. B248 (1990) 268;
P. Binetruy and M.K. Gaillard, Phys. Lett. B253
(1991) 119.

\bibitem{buscher}
T.H. Buscher, Phys. Lett. B194 (1987) 51, B201 (1988) 466.

\bibitem{rocver}
M. Ro\u{c}ek and E. Verlinde, Nucl. Phys. B373
(1992) 630.

\bibitem{givroc} A. Giveon and M.
Ro\v cek, Nucl. Phys. B380 (1992) 128.

\bibitem{aagbl}
E. {\'A}lvarez, L. {\'A}lvarez-Gaum{\'e}, J.L.F.
Barb{\'o}n and Y. Lozano, Nucl. Phys. B415 (1994) 71.

\bibitem{luis}
L. {\'A}lvarez-Gaum{\'e}, {\it Aspects of Abelian and
Non-abelian Duality},
CERN-TH-7036/93, proceedings of Trieste'93.

\bibitem{reviewkiritsis}
E. Kiritsis, Nucl. Phys. B405 (1993) 109.

\bibitem{quevedo}
X. De la Ossa and F. Quevedo, Nucl. Phys.
B403 (1993) 377.

\bibitem{givnoab}
A. Giveon and M. Ro\u{c}ek, Nucl. Phys. B421 (1994) 173.

\bibitem{r7}
M. Gasperini, R. Ricci and G. Veneziano,
Phys. Lett. B319 (1993) 438.

\bibitem{aagl}
E. {\'A}lvarez, L. {\'A}lvarez-Gaum{\'e} and Y. Lozano,
Nucl. Phys. B424 (1994) 155.

\bibitem{r11}
C. Hull and B. Spence, Phys. Lett. B232 (1989) 204.

\bibitem{quevedo2}
C.P. Burgess and F. Quevedo, Phys. Lett. B329 (1994) 457.

\bibitem{elitzur}
S. Elitzur, A. Giveon, E. Rabinovici, A. Schwimmer and G.
Veneziano,
{\it Remarks on Non-abelian Duality}, CERN-TH-7414/94,
hep-th/9409011.

\bibitem{r8}
L. {\'A}lvarez-Gaum{\'e} and E. Witten, Nucl. Phys.
B234 (1984) 269.

\bibitem{st}
A.S. Schwarz and A.A. Tseytlin, Nucl. Phys. B399 (1993) 691.

\bibitem{gilkey}
P.B. Gilkey, J. Diff. Geom. 10 (1975) 601.

\bibitem{r6}
P. Goddard, A. Kent and D. Olive, Comm. Math.
Phys. 103 (1986) 105.

\bibitem{bns}
T. Banks, D. Nemeschansky and A. Sen, Nucl. Phys. B277
(1986) 67.

\bibitem{peskin}
M.E. Peskin, {\it Introduction to String and Superstring
Theory II},
proceedings of the Theoretical Advanced Study Institute in
Elementary Particle Physics, University of California,
Santa Cruz,
1986, vol.1.

\bibitem{callan}
C. Callan, D. Friedan, E. Martinec and M. Perry,
Nucl. Phys. B262 (1985) 593.

\bibitem{horowitz}
G.T. Horowitz and D.L. Welch, Phys. Rev. Lett. 71 (1993) 328.

\bibitem{witt}
E. Witten, Phys. Rev. D44 (1991) 314.

\bibitem{gins}
P. Ginsparg and F. Quevedo, Nucl. Phys. B385 (1992) 527.

\bibitem{moore}
G. Moore, Nucl. Phys. B293 (1987) 139;
E. Alvarez and M.A.R. Osorio, Z. Phys. C44
(1989) 89.

\bibitem{nair}
V.P. Nair, A. Shapere, A. Strominger and F. Wilczek, Nucl. Phys.
B287 (1987) 402.

\bibitem{david}
F. David, Nucl. Phys. B368 (1992) 671.

\bibitem{branderva}
R. Branderberger and C. Vafa, Nucl. Phys. B316 (1989) 391.

\bibitem{ecv}
E. {\'A}lvarez, J. C{\'e}spedes and E. Verdaguer, Phys. Rev. D45
(1992) 2033; Phys. Lett. B289 (1992) 51; Phys. Lett. B304
(1993) 225.

\bibitem{venezia2}
K.A. Meissner and G. Veneziano, Phys. Lett. B267
(1991) 33.

\bibitem{aagl2}
E. {\'A}lvarez, L. {\'A}lvarez-Gaum{\'e} and Y. Lozano,
Phys. Lett. B336 (1994) 183.

\bibitem{ghandour}
G.I. Ghandour, Phys. Rev. D35 (1987) 1289;\quad
T. Curtright, {\it Differential Geometrical Methods
in Theoretical
Physics: Physics and Geometry}, ed. L.L. Chau and W. Nahm,
Plenum, New York, (1990) 279;\quad
T. Curtright and G. Ghandour, {\it Using Functional
Methods to
compute Quantum Effects in the Liouville Theory}, ed.
T. Curtright,
L. Mezincescu and R. Nepomechie, proceedings of
NATO Advanced Workshop, Coral Gables, FL, Jan 1991; and
references
therein.

\bibitem{zachos}
T. Curtright and C. Zachos, Phys. Rev. D49 (1994) 5408.

\bibitem{fj}
B.E. Fridling and A. Jevicki, Phys. Lett. B134 (1984) 70;\quad
E.S. Fradkin and A.A. Tseytlin, Ann. Phys. 162 (1985) 31.

\bibitem{WZW}
E. Witten, Comm. Math. Phys. 92 (1984) 455.


\bibitem{maldacena97} J.M. Maldacena,  hep-th/9711200.

\bibitem{klebanov97} I.R. Klebanov, {\em Nucl. Phys.} {\bf B496} (1997) 231, hep-th/9702076;
S.S. Gubser, I.R. Klebanov, and A.A. Tseytlin, {\em Nucl. Phys.} {\bf B499} (1997) 217,
hep-th/9703040; S.S. Gubser and I.R. Klebanov, {\em Phys. Lett.} 
{\bf B413} (1997) 41, hep-th/9708005; A.M. Polyakov, hep-th/9711002.

\bibitem{gubser98} S.S. Gubser, I.R. Klebanov,
and A.M. Polyakov, {\em Phys. Lett.} {\bf B428} (1998) 105,  hep-th/9802109;
E. Witten, hep-th/9802150.

\bibitem{freund} P. Freund and M. Rubin, {\em Phys. Lett.} {\bf 97B} (1980) 233;
for a review see M.J. Duff, B.E.W. Nilsson, and C.N. Pope,
{\em Phys. Rept.} {\bf 130} (1986) 1.

\bibitem{petervn} K. Pilch, P. van Nieuwenhuizen, and P.K. Townsend,
{\em Nucl. Phys.} {\bf B242} (1984) 377.

\bibitem{kim} H.J. Kim, L.J. Romans, and P. van Nieuwenhuizen,
{\em Phys. Rev.} {\bf D32} (1985) 389; M. G{\"u}naydin and N. Marcus, {\em Class. Quant. Grav.} {\bf 2} (1985) L11.

\bibitem{brink} L. Brink, J.H. Schwarz, and J. Scherk, {\em Nucl. Phys.} {\bf B121}
(1977) 77; F. Gliozzi, J. Scherk, and D. Olive, {\em Nucl. Phys.} {\bf B122} (1977) 253.

\bibitem{mandelstam} S. Mandelstam,  {\em Nucl. Phys.} {\bf B213} (1983) 149;
L. Brink, O. Lindgren, and B.E.W. Nilsson, {\em Phys. Lett.} {\bf 123B} (1983) 323;
P.S. Howe, K.S. Stelle, and P.C. West, {\em Phys. Lett.} {\bf 124B} (1983) 55;
P.S. Howe, K.S. Stelle, and P.K. Townsend, {\em Nucl. Phys.} {\bf B236} (1984) 125.

\bibitem{thooft74} G. 't Hooft, {\em Nucl. Phys.} {\bf B72} (1974) 461.

\bibitem{witten98} E. Witten, hep-th/9803131.

\bibitem{thooft93} G. 't Hooft, gr-qc/9310026; L. Susskind, {\em J. Math. Phys.} {\bf 36}
(1995) 6377, hep-th/9409089.


\bibitem{polywieg}
A.M. Polyakov and P.B. Wiegmann, Phys. Lett. B131 (1983) 121,
B141 (1984) 223.

\bibitem{bs}
I. Bars and K. Sfetsos, Mod. Phys. Lett. A7 (1992) 1091.

\bibitem{ao}
I. Antoniadis and N. Obers, Nucl. Phys. B423 (1994) 639.

\bibitem{ko}
E. Kiritsis and N. Obers, {\it A New Duality Symmetry in String
Theory}, CERN-TH 7310/94, hep-th/9406082.

radkin and A.A. Tseytlin, Ann. Phys. 162 (1985) 31.

\bibitem{WZW}
E. Witten, Comm. Math. Phys. 92 (1984) 455.

\bibitem{polywieg}
A.M. Polyakov and P.B. Wiegmann, Phys. Lett. B131 (1983) 121,
B141 (1984) 223.

\bibitem{bs}
I. Bars and K. Sfetsos, Mod. Phys. Lett. A7 (1992) 1091.

\bibitem{ao}
I. Antoniadis and N. Obers, Nucl. Phys. B423 (1994) 639.

\end{thebibliography}
\end{document}